\documentclass[12pt,letterpaper]{article}
\pdfoutput=1

\usepackage{graphicx,color}
\usepackage{amsmath,amssymb,amsfonts}
\usepackage{enumerate}
\usepackage{subfig}
\usepackage{jheppub}

\newcommand{\be}{\begin{equation}}
\newcommand{\ee}{\end{equation}}

\newcommand{\bea}{\begin{eqnarray}}
\newcommand{\eea}{\end{eqnarray}}

\newcommand{\ba}{\begin{eqnarray}}
\newcommand{\ea}{\end{eqnarray}}

\def\beq{\begin{eqnarray}}
\def\eeq{\end{eqnarray}}

\begin{document}

\begin{flushright}
HIP-2019-10/TH
\end{flushright}

\begin{center}

\centerline{\Large {\bf Notes on entanglement wedge cross sections}}

\vspace{8mm}

\renewcommand\thefootnote{\mbox{$\fnsymbol{footnote}$}}
Niko Jokela${}^{1,2}$\footnote{niko.jokela@helsinki.fi}
and Arttu P\"onni${}^{1,2}$\footnote{arttu.ponni@helsinki.fi}

\vspace{6mm}

${}^1${\small \sl Department of Physics} and ${}^2${\small \sl Helsinki Institute of Physics} \\
{\small \sl P.O.Box 64, FIN-00014 University of Helsinki, Finland}

\end{center}

\vspace{6mm}

\setcounter{footnote}{0}
\renewcommand\thefootnote{\mbox{\arabic{footnote}}}

\begin{abstract}
\noindent

We consider the holographic candidate for the entanglement of purification $E_P$, given by the minimal cross sectional area of an entanglement wedge $E_W$. The $E_P$ is generally very complicated quantity to obtain in field theories, thus to establish the conjectured relationship one needs to test if $E_W$ and $E_P$ share common features. In this paper the entangling regions we consider are slabs, concentric spheres, and creases in field theories in Minkowski space. The latter two can be mapped to regions in field theories defined on spheres, thus corresponding to entangled caps and orange slices, respectively. We work in general dimensions and for slabs  we also consider field theories at finite temperature and confining theories. We find that $E_W$ is neither a monotonic nor continuous function of a scale. We also study a full ten-dimensional string theory geometry dual to a non-trivial RG flow of a three-dimensional Chern-Simons matter theory coupled to fundamentals. We show that also in this case $E_W$ behaves non-trivially,  which if connected to $E_P$,  lends further support that the system can undergo purification simply by expansion or reduction in scale.

\end{abstract}

\newpage
\tableofcontents

\section{Introduction}

Much attention has been paid to the quantum entanglement entropy for pure states.  The entanglement entropy is easy to define while is typically difficult to compute. In AdS/CFT the gravity dual of this quantity is the Ryu-Takayanagi formula \cite{Ryu:2006bv}, a simple computation of a minimal area of a bulk surface anchored on a boundary region of interest. The RT formula has passed several nontrivial tests and it is interesting to ask if one can extend the holographic vantage point to computing other even more difficult information theoretic quantities for interacting QFTs.

In recent years we have witnessed several attempts in understanding the entanglement entropy of mixed states using holographic methods. In particular, the entanglement of purification $E_P$ \cite{EoP} and negativity ${\cal E}$ \cite{Vidal:2002zz} have gathered lots of recent attention. A candidate holographic counterpart is the entanglement wedge cross section $E_W$ \cite{Takayanagi:2017knl}, the minimal cross section of the entanglement wedge. Since sparse examples exist for which this can be computed on the field theory side, it is currently far from clear in which cases the conjectured relationship between $E_W$ and $E_P$ or ${\cal E}$ holds.

One of the complications to interpret $E_W$ as the entanglement of purification or negativity, is that $E_W$ is UV finite by construction while the entanglement measures are divergent and subject to regularization, thus leading to subtleties. Nevertheless, these entanglement measures quantify certain correlations between subsystems and their {\emph{behavior}} is known. For example, they should be continuous and monotonic under local operations \cite{EoP}, while they need not be convex \cite{Plenio:2005cwa}.
Therefore, one can similarly investigate the properties of the entanglement wedge cross section and test whether it satisfies the same inequalities as the correlation measures.
An inequality of this sort $E_P(\rho_{AB}) \geq \frac{1}{2}I(A,B) $ represents a particularly clean example. By replacing for the entanglement wedge cross section on the left-hand-side brings us to:
\be\label{eq:ineq}
 E_W(\rho_{AB}) \geq \frac{1}{2}I(A,B)
\ee
and thus instructs us to test if $E_W$ always at least exceeds half the mutual information $I(A,B)$, which is also a UV finite quantity by construction. This inequality has been proven in holography \cite{Takayanagi:2017knl}. Inequalities for several boundary regions for entanglement of purification are also known (see, {\emph{e.g.}}, \cite{Nguyen:2017yqw}), but such analogous inequalities for $E_W$ have not been exhaustingly tested (see, however, \cite{Du:2019emy}). In all of the cases we study in this paper we found that the inequality (\ref{eq:ineq}) is satisfied.
However, this is more exciting than it sounds: $I(A,B)$ can have very non-trivial features if conformal symmetry is broken \cite{Balasubramanian:2018qqx}, thus drawing attention to $E_W$ in same geometries.
Moreover, the advances put forward in this paper enable one to test if $E_W$ could also satisfy further inequalities known to hold for $E_P$ involving three or more regions. We plan to return to these interesting questions in the future.

Let us remark that in \cite{Chaturvedi:2016rcn,Jain:2017aqk,Jain:2017xsu} it was suggested that in holographic framework the mutual information, multiplied by a constant, could be a dual to the entanglement negativity ${\cal E}$. This was revised by a suggestion \cite{Kudler-Flam:2018qjo} that the entanglement wedge cross section $E_W$, supplemented by the backreaction of the hanging minimal (cosmological) surfaces to the geometry, is actually a gravity dual of entanglement negativity. While lots of important work \cite{Takayanagi:2017knl,Chaturvedi:2016rcn,Jain:2017aqk,Jain:2017xsu,Nguyen:2017yqw,Hirai:2018jwy,Espindola:2018ozt,Kudler-Flam:2018qjo,Tamaoka:2018ned,Bao:2018fso,Bao:2018zab,Kang:2018xqy,Agon:2018lwq,Yang:2018gfq,Bao:2018pvs,Caputa:2018xuf,Guo:2019azy,Bhattacharyya:2019tsi,Bao:2019fpq,Liu:2019qje,Ghodrati:2019hnn,Kudler-Flam:2019oru,BabaeiVelni:2019pkw,Du:2019emy} has been carried out and most obvious inequalities tested, we feel that it is premature to identify $E_W$ one way or the other. In particular, as we will discuss in the following, $E_W$ can be discontinuous and non-monotonous, unlike $E_P$ or ${\cal E}$ in gauge theories at finite $N$.

In this paper we do not resolve the interpretational issues of the entanglement wedge cross section. However, we will make important headway towards reaching this goal by making several non-trivial technical advances in computing $E_W$ is various settings. We note, that almost all explicit computations of $E_W$ have been performed in asymptotically $AdS_3$ in global coordinates, where the boundary regions $A$ and $B$ are segments of a circle so that the region anchoring the entanglement wedge is the rest of the boundary spacetime.
One of our results is to generalize the computation of $E_W$ to any number of spacetime dimensions, where the boundary regions $A,B$ generalize to caps of hyperspheres. We find this computation most straightforward by first solving a related problem in Poincar\'e patch and then mapping the results to global coordinates, as detailed in Appendix~\ref{app:mapping}.

Our primary goal is to extend the notion $E_W$ and computations thereof to Poincar\'e coordinates. We will consider several different bulk spacetimes, starting with $AdS$ in arbitrary dimension. One important result is that we are able to obtain $E_W$ analytically for many different entangling surfaces. We have organized the paper in different sections according to boundary entangling geometries as follows: slabs in Sec.~\ref{sec:strips}, spheres in Sec.~\ref{sec:annulus}, and creases in Sec.~\ref{sec:slice}. In the case of slabs, in addition to pure AdS in arbitrary dimension, we furthermore work out $E_W$ in the presence of a black brane, {\emph{i.e.}}, duals of conformal field theories at finite temperature. In addition to this, we also consider general confining geometries, where the phase diagram for the two entangling surfaces is quite rich. As an illustration of a novel effect in this scenario we will portray $E_W$ in a configuration where it can suddenly jump upwards at larger distances. 

As an extension to a more complicated situation, we also display some results in Sec.~\ref{sec:ABJM} in the full 10d string theory background, which represents the gravity dual to a Chern-Simons matter theory flowing between two different RG fixed points. We find that the entanglement wedge cross section is not monotonic under RG flow, or equivalently, at different length scales in the following sense. Given the field theory at zero temperature and considering two strip regions which initially share mutual information, the system can undergo a purification just due expansion or reduction. In terms of entanglement wedge cross section this corresponds to vanishing values both at large and small distances, but not at intermediate length scales. This is in sharp contrast to conformal field theories, where the system either shares mutual information or not at any length scale. If the number of fundamentals is not too large, our results assume analytic forms (see Appendix~\ref{app:abjm}), yielding good geometric understanding behind this phenomenon.

\vspace{1cm}

{\bf{Note added:}} While this paper was in the final stages \cite{Liu:2019qje,Ghodrati:2019hnn,BabaeiVelni:2019pkw} appeared, which have partial overlap with our results.

\newpage

\section{Definitions}

Consider two subsystems of the boundary, $A$ and $B$ with no non-zero overlap. Let $\Gamma_{AB}^{min}$ denote the minimal RT-surface associated with the union $AB$. The {\it entanglement wedge} $M_{AB}$ is the bulk region whose boundary is
\begin{align} \label{eq:wedge}
   \partial M_{AB} = A \cup B \cup \Gamma_{AB}^{min} \ .
\end{align}
Note that if $A$ and $B$ are small enough or there is enough separation, $M_{AB}$ will become disconnected. Also note that actually the entanglement wedge is the bulk codimension-0 region which is the domain of dependence of $M_{AB}$ but since we are working on time slices of static backgrounds this distinction is not relevant at this moment.

Now split $\partial M_{AB}$ into two pieces
\begin{align} \label{eq:wedge_split}
   \partial M_{AB} = \widetilde{\Gamma}_A \cup \widetilde{\Gamma}_B
\end{align}
such that $A \subset \widetilde{\Gamma}_A$ and $B \subset \widetilde{\Gamma}_B$.

Then we search for a minimal surface $\Sigma_{AB}^{min}$ subject to
\begin{align}
   (i) \quad & \partial \Sigma_{AB}^{min} = \partial \widetilde{\Gamma}_A = \partial \widetilde{\Gamma}_B \\
   (ii) \quad & \Sigma_{AB}^{min} \text{ is homologous to } \widetilde{\Gamma}_A \ .
\end{align}
There is an infinite set of possible splits and thus infinite set of possible $\Sigma_{AB}^{min}$. The {\it entanglement wedge cross section} is given by the volume of $\Sigma_{AB}^{min}$ minimized over all possible splits of the entanglement wedge
\begin{align}
   E_W (\rho_{AB}) = \underset{\widetilde{\Gamma}_A \subset \partial M_{AB}}{min} \left( \frac{A(\Sigma_{AB}^{min})}{4 G_N} \right) \ .
\end{align}
In other words, $E_W (\rho_{AB})$ is the minimal area surface that splits the entanglement wedge $M_{AB}$ into two regions, one for $A$ and another for $B$.

\section{Slabs}\label{sec:strips}

\begin{figure}
   \centering
   \includegraphics[width=1.0\textwidth]{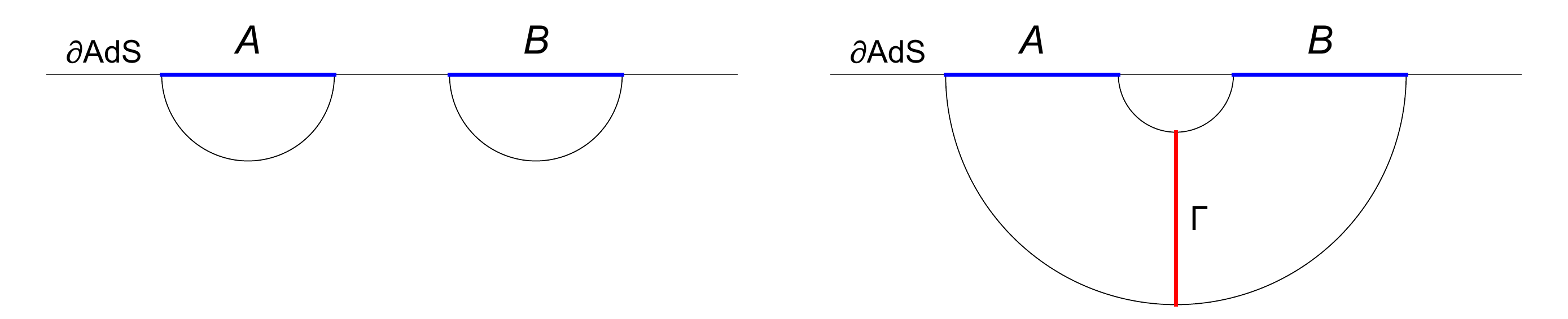}
   \caption{Sketch of the different entanglement entropy phases available for parallel strip configurations of width $l$ and separated by a distance $s$. The disconnected phase (left) has zero entanglement wedge cross section. The connected phase (right) does have a non-zero $E_W$, given by the area of $\Gamma$, the surface with the minimal area out of the infinite set of surfaces $\Sigma_{AB}^\text{min}$.}
   \label{fig:poincare_wedge_sketch}
\end{figure}

In this section we will start computing the entanglement wedge cross sections in various bulk geometries. The configurations that we will consider are\footnote{We work in dimensions $d>2$.} pure $AdS_{d+1}$, planar $AdS_{d+1}$ black brane, $AdS_{d+1}$ soliton, massless ABJM at finite temperature, and massive ABJM at zero temperature. For simplicity, we will only consider symmetric configurations of strips, that is, two parallel strips with equal widths $l$, separated by a distance $s$. We have sketched this configuration in Fig.~\ref{fig:poincare_wedge_sketch}. All the computations presented in this section can be easily extended to non-symmetric situations.

\subsection{Pure $AdS_{d+1}$}\label{sec:strips_pure}

We start with the case of pure $AdS_{d+1}$ in the Poincar\'e patch. The bulk metric reads
\begin{align}
   g = \frac{L_{AdS}^2}{z^2} \left( -dt^2+ dz^2+ dx^2+ d\vec{x}_{d-2}^2 \right) \ .
\end{align}
The boundary is at $z=0$ and all strip boundaries lie on lines of constant $x$.

In pure $AdS_{d+1}$, the dominant phase is determined by a single number, the ratio between the separation and the width of the strips $s/l$. When the strips are close enough to each other, the system is in the connected phase. Otherwise the strips are too far apart and the corresponding RT-surfaces disconnect. The $s/l$ at which the transition happens is given by the real, positive root of \cite{Ben-Ami:2014gsa}
\begin{align}
   \frac{1}{(2+s/l)^{d-2}} + \frac{1}{(s/l)^{d-2}} = 2  \label{eq:crit_s_per_l} \ .
\end{align}
Assuming $s/l$ is such that the system is in its connected phase, the entanglement wedge cross section can be calculated as follows. The bulk turning point of a strip of width $l$ is
\be\label{eq:AdSl}
   z_*(l) = \frac{\Gamma\left(\frac{1}{2(d-1)}\right)}{2\sqrt\pi\Gamma\left(\frac{d}{2(d-1)}\right)} l \ .
\ee
The surface which splits the entanglement wedge into two parts associated with $A$ and $B$ and does so with minimal area, $\Gamma$, can be identified by symmetry to be a vertical, flat surface which splits the entanglement wedge at its symmetry axis. The induced metric on $\Gamma$, see Fig.~\ref{fig:poincare_wedge_sketch}, is
\begin{align}
   g_\Gamma = \frac{L_{AdS}^2}{z^2}(dz^2 + d\vec{x}_{d-2}^2) \ ,
\end{align}
where $d\vec{x}_{d-2}^2$ contains the transverse directions. The entanglement wedge cross section then reads
\bea
   E_W & = & \frac{A(\Gamma)}{4 G_N^{(d+1)}} = \frac{1}{4 G_N^{(d+1)}} \int_\Gamma vol_\Gamma \nonumber\\
   & = & \frac{V L_{AdS}^{d-1}}{4 G_N^{(d+1)} (d-2)} \left( \frac{1}{z_*(s)^{d-2}} - \frac{1}{z_*(2l+s)^{d-2}} \right) \label{eq:strip_EW_zs} \\
   & = & \frac{V L_{AdS}^{d-1}}{4 G_N^{(d+1)} (d-2)} \frac{2^{d-2}\pi^\frac{d-2}{2}\Gamma\left(\frac{d}{2(d-1)}\right)^{d-2}}{\Gamma\left(\frac{1}{2(d-1)}\right)^{d-2}} \left( \frac{1}{s^{d-2}} - \frac{1}{(2l+s)^{d-2}} \right) \ ,
\eea
where $V=L_y L_z \dots$ is the volume of transverse directions.

\begin{figure}
   \centering
   \includegraphics[width=0.6\textwidth]{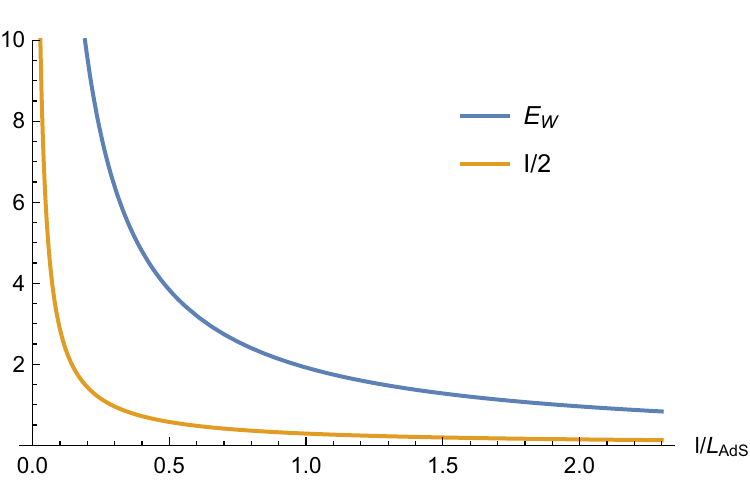}
   \caption{Comparison of $E_W$ and $I/2$ in pure $AdS_4$. Entanglement regions are parallel strips with width $l$ and separation $s$ fixed such that $s/l=0.4$. In both quantities we have omitted a prefactor of $\frac{V L_{AdS}^2}{4 G_N^{(4)}}$.}
   \label{fig:ads_poincare}
\end{figure}

In Fig.~\ref{fig:ads_poincare} we have depicted $E_W$ and $I/2$ for parallel strips of equal widths in pure $AdS_4$. The ratio of strip separation and width ($s/l$) is fixed to a constant value. As long as the separation is less than that of (conjugate) golden ratio $s/l < \frac{\sqrt{5}-1}{2}$, different values of $s/l$ yield qualitatively similar curves: monotonically decreasing curves with $E_W > I/2$. If separation is greater than the critical value $s/l=\frac{\sqrt{5}-1}{2}$, the RT-surface for $S(AB)$ disconnects resulting in $E_W=I=0$.

\subsection{$AdS_{d+1}$ black brane}
Let us now generalize the previous case in field theories at finite temperature. This corresponds to focusing on $AdS$-Schwarzschild geometries. The metric reads
\begin{align}
   g = \frac{L_{AdS}^2}{z^2} \left( -b(z)dt^2+ \frac{dz^2}{b(z)}+ dx^2 + d\vec{x}_{d-2}^2 \right) \ ,
\end{align}
where the blackening factor is $b(z)=1-\frac{z^d}{z_h^d}$.

Symmetry of the strip configuration and bulk metric still impose that, assuming that the entanglement wedge is connected, the entanglement wedge cross section is given by the area of a constant-$x$ hypersurface, $\Gamma$, located in the middle of the strips. This time the induced metric on $\Gamma$ is
\begin{align}
   g_\Gamma = \frac{L_{AdS}^2}{z^2} \left( \frac{dz^2}{b(z)} + dx^2+ d\vec{x}_{d-2}^2 \right)\ .
\end{align}
The area of $\Gamma$, and thus $E_W$ is determined by the $z$-coordinates of $\partial\Gamma$, $z_*(s)$, and $z_*(2l+s)$, which are in turn determined by $l$ and $s$. The entanglement wedge cross section, in terms of $z_*(s)$ and $z_*(2l+s)$ is
\begin{align}
   E_W &= \frac{1}{4 G_N^{(d+1)}} \int vol_\Gamma \\
   &= \frac{V L_{AdS}^{d-1} z_h^{2-d}}{4 G_N^{(d+1)}(d-2)} \Bigg( \left(\frac{z}{z_h}\right)^{2-d} {}_2 F_1 \left( \frac{1}{2}, \frac{2-d}{d}, \frac{2}{d}, \frac{z^d}{z_h^d} \right) \Bigg)_{z_*(2l+s)}^{z_*(s)} \ ,
\end{align}
where $z_*(l)$ is the function giving the strip turning point for a given width $l$. The width of a strip can be expressed as a series \cite{Fischler:2012ca,Erdmenger:2017pfh}
\begin{align}
   l = 2\sqrt\pi z_* \sum_{m=0}^\infty \frac{1}{m!(1+m d)} \left( \frac{1}{2} \right)_m \frac{\Gamma\left( \frac{d(m+1)}{2(d-1)} \right)}{\Gamma\left( \frac{1+m d}{2(d-1)} \right)} \left(\frac{z_*}{z_h}\right)^{md} \ ,
\end{align}
where $(a)_n$ is the Pochhammer symbol. Notice that the zero temperature limit $z_h\to 0$ picks the first term in the series, thus reducing to (\ref{eq:AdSl}). This series can be reverted for $z_*(l)$.

\begin{figure}
   \centering
   \includegraphics[width=0.8\textwidth]{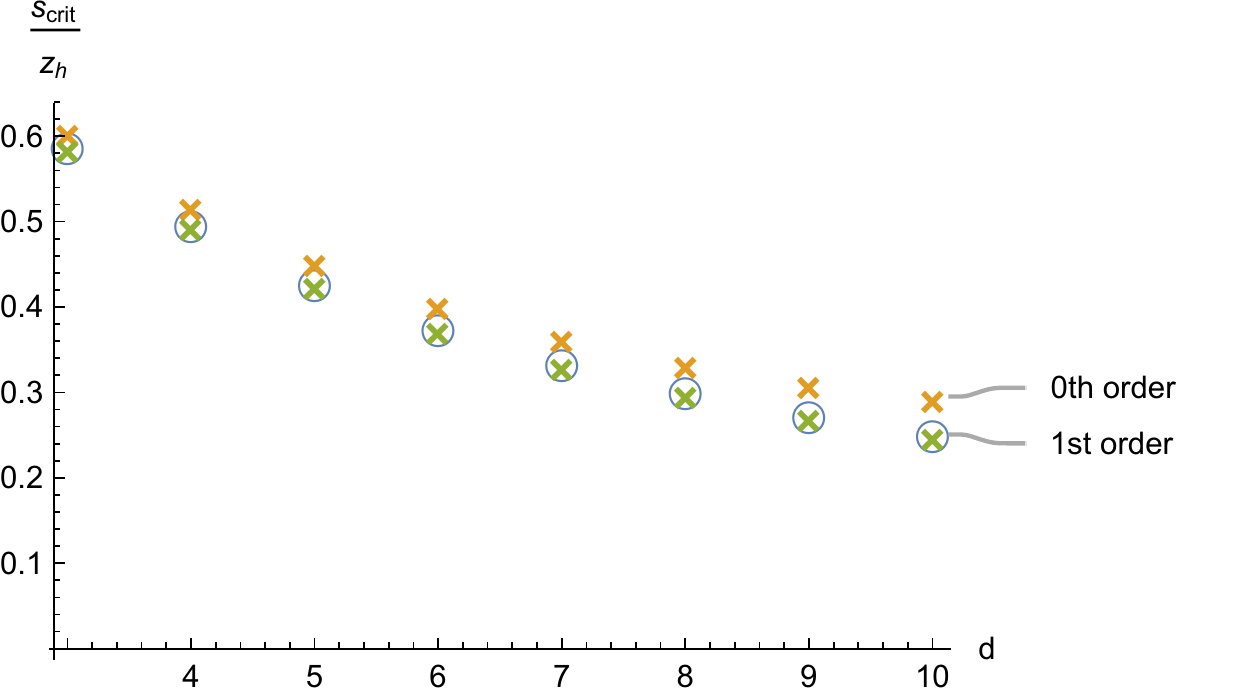}
   \caption{Critical separation between parallel strips in $AdS_d-BH$ for $d=2,\dots,10$. Circles denote numerical results. Green and orange crosses correspond to analytical approximations (\ref{eq:crit_s_1}) and (\ref{eq:crit_s_2}), respectively.}
   \label{fig:crit_s}
\end{figure}

One important difference between this black hole background and the pure $AdS_{d+1}$ of previous section is that there exists a critical separation $s_c$ such that if $s>s_c$, no connected phase exists for any $l$ \cite{Yang:2018gfq}. This also means that $E_W=0$ if $s>s_c$. Now we work out an analytical approximation for $s_c$. The phase transition between connected and disconnected phases happens when the corresponding mutual information vanishes
\bea
   I(l,s) & = & S_A+S_B-S_{AB} \nonumber \\
   & = & 2 S(l) - S(2l+s) - S(s) = 0 \ \label{eq:mutual_zero} \ .
\eea
Since we should consider the limit $l/z_h\to\infty$, all terms involving $l$ are well represented by an IR-formula for the entanglement entropy. We will calculate the remaining term, $S(s)$, in an UV-expansion. In the UV, we have
\bea
   S(l) & = &  \frac{L_{AdS}^{d-1} V}{2(d-2) G_N^{(d+1)}} \left( \frac{1}{\epsilon^{d-2}} - \frac{2^{d-2}\pi^\frac{d-1}{2}\Gamma\left(\frac{d}{2(d-1)}\right)^{d-1}}{\Gamma\left(\frac{1}{2(d-1)}\right)^{d-1} l^{d-2}} \right) \nonumber \\
   & & + \frac{L_{AdS}^{d-1}V z_h^{-d}}{32(d+1)G_N^{(d+1)}} \frac{\Gamma\left(\frac{1}{2(d-1)}\right)^2\Gamma\left(\frac{1}{d-1}\right)}{\sqrt\pi\Gamma\left(\frac{1}{2}+\frac{1}{d-1}\right)\Gamma\left(\frac{d}{2(d-1)}\right)^2} l^2 \ , \label{eq:S_bh_pp_UV}
\eea
where $\epsilon$ stands for UV cut-off. In the IR, on the other hand \cite{Fischler:2012ca,Erdmenger:2017pfh},
\be
   S(l) = \frac{L_{AdS}^{d-1} V}{2(d-2)G_N^{(d+1)}} \frac{1}{\epsilon^{d-2}} + \frac{L_{AdS}^{d-1}V}{4G_N^{(d+1)}z_h^{(d-2)}} \Bigg( \frac{l}{z_h} + 2\mathcal{C}(z_h) \Bigg) \ , \label{eq:S_bh_pp_IR}
\ee
where $\mathcal{C}(z_h)$ is an ${\cal O}(1)$ numerical constant whose values can be found in, {\emph{e.g.}}, \cite{Erdmenger:2017pfh}.
By considering only the leading order UV contribution to $S(s)$, (\ref{eq:mutual_zero}) is equivalent to
\begin{align}
   \left(2\mathcal{C}(z_h)-\frac{s}{z_h}\right) + \frac{2^{d-1}\pi^\frac{d-1}{2}\Gamma\left(\frac{d}{2(d-1)}\right)^{d-1}}{(d-2)\Gamma\left(\frac{1}{2(d-1)}\right)^{d-1}} \left(\frac{s}{z_h}\right)^{2-d} = 0 \ . \label{eq:crit_s_1}
\end{align}
This gives a first approximation for the critical $s/z_h$. We can systematically improve the approximation by taking more UV-terms in the calculation. Taking the subleading UV-term in (\ref{eq:S_bh_pp_UV}) into account, we find
\begin{gather}
   \left(2\mathcal{C}(z_h)-\frac{s}{z_h}\right) + \frac{2^{d-1}\pi^\frac{d-1}{2}\Gamma\left(\frac{d}{2(d-1)}\right)^{d-1}}{(d-2)\Gamma\left(\frac{1}{2(d-1)}\right)^{d-1}} \left(\frac{s}{z_h}\right)^{2-d} \nonumber \\
   - \frac{1}{8(d+1)} \frac{\Gamma\left(\frac{1}{2(d-1)}\right)^2\Gamma\left(\frac{1}{d-1}\right)}{\sqrt\pi \Gamma\left(\frac{1}{2}+\frac{1}{d-1}\right)\Gamma\left(\frac{d}{2(d-1)}\right)^2} \left(\frac{s}{z_h}\right)^2 = 0 \ . \label{eq:crit_s_2}
\end{gather}
Already at this level of approximation, the agreement with numerics is remarkably good. In Fig.~\ref{fig:crit_s} we have compared the numerical results with both the leading and to next-to-leading order approximations.

\begin{figure}
   \centering
   \includegraphics[width=0.6\textwidth]{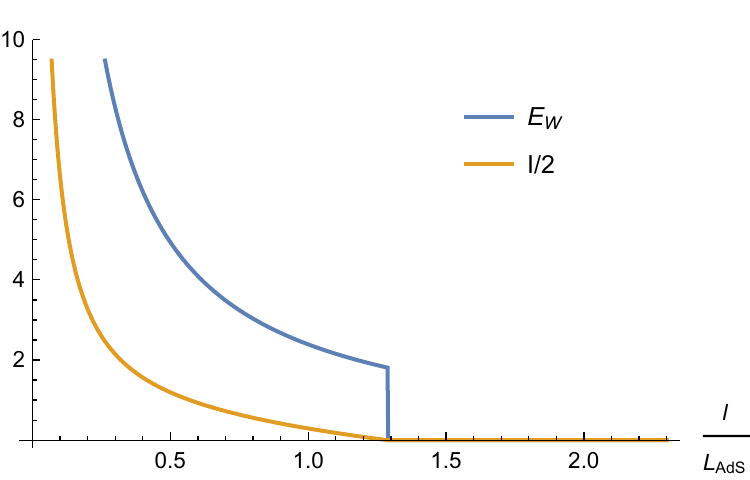}
   \caption{Comparison of $E_W$ and $I/2$ in an asymptotically $AdS_4$ black brane geometry. We have again fixed $s/l=0.4$ and omitted a common prefactor of $V L_{AdS}^2 z_h/4 G_N^{(4)}$. A qualitative difference to the pure $AdS_4$ (Fig.~\ref{fig:ads_poincare}) is the appearance of a phase transition of $S(AB)$ to the disconnected phase marked by the point where both $E_W$ and $I/2$ become zero.}
   \label{fig:ads_bh_poincare}
\end{figure}

In Fig.~\ref{fig:ads_bh_poincare} we show the same quantities in the same setup that in Fig.~\ref{fig:ads_poincare} except in an asymptotically $AdS_4$ black brane geometry. The different quantities $E_W$ and $I$ become zero when the RT-surface prefers the disconnected phase. They do this in a different way though: $I(A,B)$ becomes zero continuously since the phase transition point is determined by $I(A,B)=S(A)+S(B)-S(AB)=0$. The terms are continuous and thus $I(A,B)$ is also. The entanglement wedge cross section $E_W$, on the other hand, jumps discontinuously at the phase transition because on the other side the entanglement wedge is disconnected and on the side where it is connected it has a finite cross section as can be easily seen in the sketch in Fig.~\ref{fig:poincare_wedge_sketch}: there is no need for $\Gamma$ to pinch to zero at the phase transition.

Having discussed the simple $AdS$ and $AdS$-BH geometries in various dimensions, now we will turn to cases that have a non-trivial length scale associated with them. First, we will discuss confining geometries, again, in all dimensions at once. Later, we will focus on a specific top-down construction that is dual to a Chern-Simons gauge theory coupled with fundamental matter. The important difference relative to previous ones is that in both of these examples the RT-surfaces are sensitive to the underlying scale, leading to drastic effects for the information quantities. For example, we will find that, in general, neither the mutual information $I$ nor the entanglement wedge cross section $E_W$ are monotonic, but show structure at the underlying length scales.

\subsection{Confining backgrounds}

We will continue computing $E_W$ and $I$ for two-strip configurations in confining geometries; for studies of holographic entanglement entropy in this context, see, {\emph{e.g.}, \cite{Nishioka:2006gr,Klebanov:2007ws,Nishioka:2009un,Ben-Ami:2014gsa,Kol:2014nqa,Georgiou:2015pia}}. A family of confining geometries are easily obtained from the AdS-Schwarzschild metric via double Wick rotation. Note, however, as in the previous cases, we do not consider non-trivial dilaton profiles. The metrics we consider are
\begin{align}\label{eq:confmetric}
   g = \frac{L_{AdS}^2}{z^2} \left( -dt^2+ \frac{dz^2}{b(z)} + dx^2 + d\vec{x}_{d-3}^2 + b(z) dx_\text{circle}^2\right) \ ,
\end{align}
where $b(z)=1-\frac{z^d}{z_h^d}$ and the radius of the circle is related to $z_h$.
For example, the case $d=4$ is the AdS-soliton solution \cite{Witten:1998zw}, whose gauge theory dual is the ${\cal{N}}=4$ super Yang-Mills theory on $\mathbb{R}^{1,2}\times S^1$.

Induced metric on $\Gamma$, a surface with $t=\text{const}, x=\text{const}$, is
\begin{align}
   g_\Gamma = \frac{L_{AdS}^2}{z^2} \left( \frac{dz^2}{b(z)} + d\vec{x}_{d-3}^2 + b(z) dx_\text{circle}^2\right) \ .
\end{align}
The entanglement wedge cross section is
\begin{align}
   E_W &= \frac{V L_{AdS}^{d-1}}{4 G_N^{(d+1)}} \int_{z_*(s)}^{z_*(2l+s)} \frac{dz}{z^{d-1}} \\
   &= \frac{V L_{AdS}^{d-1} z_h^{2-d}}{4 G_N^{(d+1)} (d-2)} \left[ \left(\frac{z_h}{z_*(2l+s)}\right)^{d-2} - \left(\frac{z_h}{z_*(s)}\right)^{d-2} \right] \ . \label{eq:strip_EW_zs_confining}
\end{align}
One can notice that upon calculating the determinant of $g_\Gamma$, the factors of $b(z)$ cancel out leaving behind the same expression as we encountered in Sec. \ref{sec:strips_pure} for pure $AdS_{d+1}$. In terms of strip widths $E_W$ still differs from the pure $AdS_{d+1}$ result since in this confining geometry, the function $z_*(l)$, is quite different.

For a single strip, the width/entropy integrals are
\begin{align}
   l(z_*) &= 2 \int \left( \frac{z}{z_s} \right)^{d-1} \frac{\sqrt{b(z_*)}}{b(z)} \frac{1}{\sqrt{1-\left(\frac{z}{z_*}\right)^{2(d-1)} \frac{b(z_*)}{b(z)}}} dz \\
   S(z_*) &= \frac{L_{AdS}^{d-1} V}{2(d-2)G_N^{(d+1)}} \frac{1}{\epsilon^{d-2}} \nonumber \\
   &+ \frac{L_{AdS}^{d-1}V}{2 G_N^{(d+1)}} \int_0^{z_*} \Bigg(\frac{1}{z^{d-1}} \frac{1}{\sqrt{1-\left(\frac{z}{z_*}\right)^{2(d-1)}\frac{b(z_*)}{b(z)}}} - \frac{1}{z^{d-1}} \Bigg) dz - \frac{1}{(d-2) z_*^{d-2}} \ .
\end{align}
These integrals are needed in order to find the function $z_*(l)$ appearing in eq. (\ref{eq:strip_EW_zs_confining}) and for finding the phase of the bipartite system.

\begin{figure}
   \centering
   \hspace*{-2cm}
   \includegraphics[width=1.3\textwidth,clip,trim={2.5cm 0 0.5cm 4cm}]{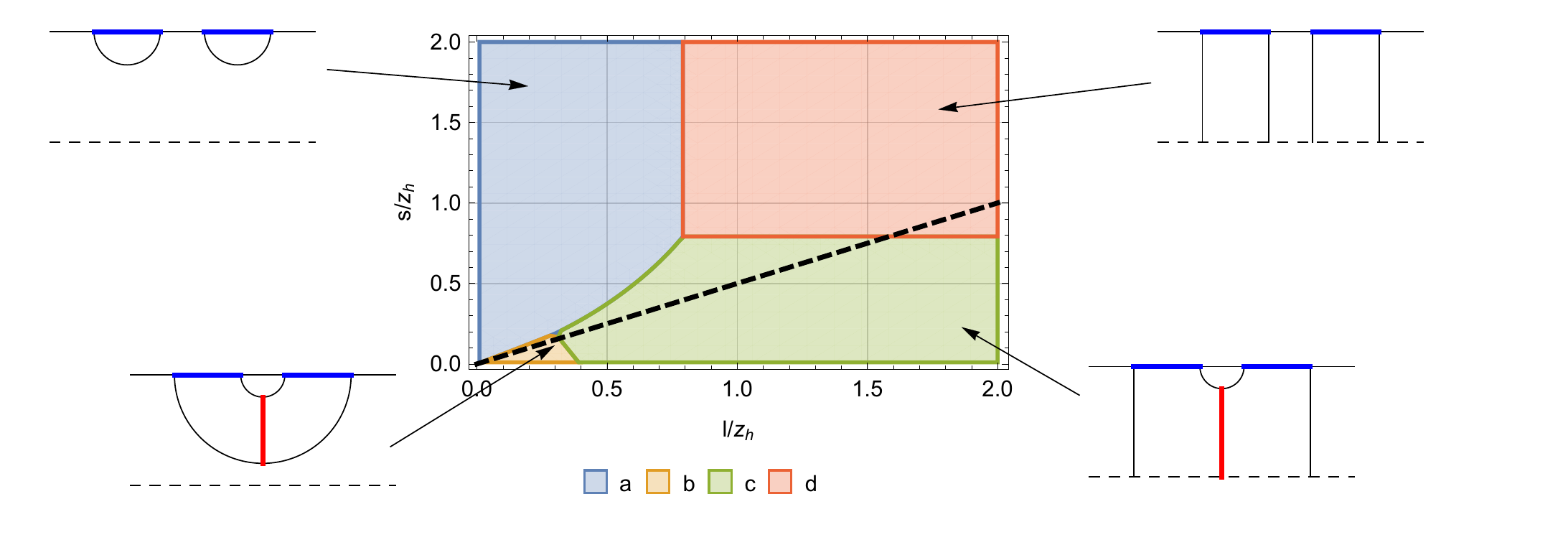}
   \caption{Phase diagram for a bipartite system composed out of two parallel, infinitely long strips for $d=3$ \cite{Ben-Ami:2014gsa}. There are four possible bulk surfaces for $S(AB)$, where $A$ and $B$ have widths $l$ and are separated by $s$. Only two of these give rise to a connected entanglement wedge and thus a non-zero $E_W$, denoted by a red line in the sketches. The black dashed line shows the parameter space of Fig.~\ref{fig:confining_EW}.}
   \label{fig:poincare_wedge_sketch_confining}
\end{figure}

For a bipartite entanglement region comprised of two parallel strips, there exists four possible phases of $S(AB)$. These four phases, that we call $a,b,c,d$, are sketched in Fig.~\ref{fig:poincare_wedge_sketch_confining} along with a phase diagram for a system where strips have widths $l$ and are separated by a distance $s$ \cite{Ben-Ami:2014gsa}.\footnote{Interestingly, a similar four-zone phase diagram occurs also in an anisotropic system \cite{Jokela:2019tsb}.} Notice that the slope of the phase diagram between the phases $a$ and $b$, close to the UV, is given by the (conjugate) golden ratio $\varphi^{-1}=(\sqrt 5-1)/2$ in $d=3$ \cite{Balasubramanian:2018qqx}; in general $d$ via the root of (\ref{eq:crit_s_per_l}) \cite{Ben-Ami:2014gsa}.

\begin{figure}
   \centering
   \includegraphics[width=0.6\textwidth]{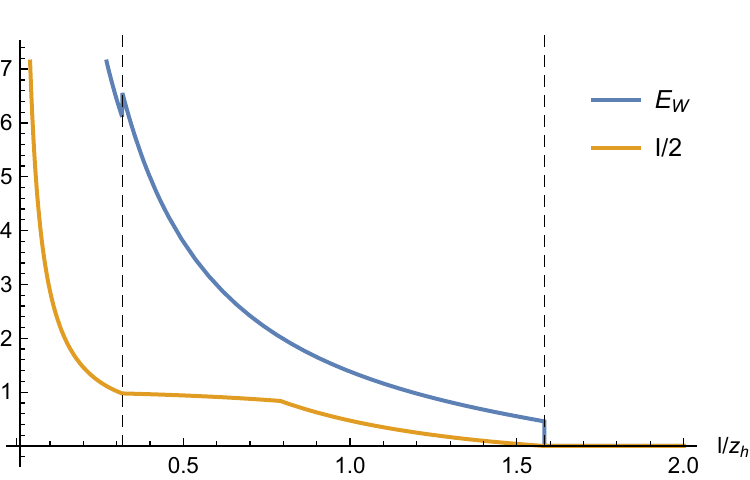}
   \caption{Entanglement wedge cross section for parallel strips in $d=3$, where we have fixed $s/l=0.5$. Plotting $E_W$ with this $s/l$ corresponds to moving along the black dashed line in Fig.~\ref{fig:poincare_wedge_sketch_confining}. Vertical black dashed lines correspond to phase transitions in $S(AB)$. The leftmost transition happens when $2l+s=l_\text{crit}^{(d)}$ and is interesting because in this situation $E_W$ jumps discontinuously to a larger positive value when strips increase in size. The rightmost transition corresponds to $s=l_\text{crit}^{(d)}$, where the entanglement wedge becomes disconnected. Note that the mutual information has an additional phase transition corresponding to the connected-disconnected transition of $S(A)$ which occurs at $l=l_\text{crit}^{(d)}$ and is responsible for the cusp of $I(A,B)$.}
   \label{fig:confining_EW}
\end{figure}

In order to show results for the information measures, we will pick units in terms of $z_h$. In Fig.~\ref{fig:confining_EW} we depict $E_W$ as a function of $l/z_h$ for fixed $s/l$. The ratio $s/l$ is chosen such that we will pass through both phases with $E_W>0$, indicated as a thick dashed black line in Fig.~\ref{fig:poincare_wedge_sketch_confining}. There are non-trivial features appearing each time a phase transition between different phases for the strips occur. In Fig.~\ref{fig:confining_EW} we show results for the mutual information, again bounded from above by $E_W$ as expected. In this confining background also the mutual information has some non-trivial features.  Close to the IR, the strips are in a ``disconnected'' phase, {\emph{i.e.}}, they go all the way to the tip of the cigar in which point they connect due to pinching of the circle direction. For large values of the strip widths then both $E_W$ and $I/2$ vanish, with the former discontinuously. In the UV region, on the other hand, both $E_W$ and $I/2$ follow the trend of a typical AdS background, being also in the $b$ configuration. In between the two extremes we have a phase transition $b\to c$ intrinsic to a confining background. For the $E_W$ this shows up as a jump upwards: all of the sudden the wedge cross section grows, as is clear from the sketches in Fig.~\ref{fig:poincare_wedge_sketch_confining}. The mutual information is continuous across this phase transition but has a cusp. Later, but when $S(AB)$ is still in the $c$ phase, the $I/2$ has another cusp. This is because
the individual terms $S(A)$ and $S(B)$ undergo a phase transition to the ``disconnected'' phases.

\subsection{Chern-Simons theory coupled with fundamentals in $2+1$ dimensions}\label{sec:ABJM}

Let us now focus on a top-down construction and consider the gravity dual to a known quantum field theory. We will consider the ABJM Chern-Simons matter theory \cite{Aharony:2008ug} coupled with fundamental matter in $2+1$ dimensions in the Veneziano limit \cite{Conde:2011sw}. The entanglement entropies for this system have been considered in previous works \cite{Bea:2013jxa,Balasubramanian:2018qqx} and here we will comment on results for the entanglement wedge cross section.
The background with fundamental matter we consider here are of two-fold:\footnote{Generalizations of the current work to the cases with parity breaking \cite{Bea:2014yda} or to noncommutativity at the UV \cite{Bea:2017iqt} could also lead to potentially interesting surprises.} either masses of the fundamentals are zero \cite{Jokela:2012dw} or they are massive but at vanishing temperature \cite{Bea:2013jxa}.

In the former case, {\emph{i.e.}}, for massless fundamental matter coupled to ABJM Chern-Simons theory, the results of the preceding subsections go through almost unchanged. This is because the inclusion of fundamental matter keeps the field theory conformal, while the number of degrees of freedom increase, a fact which is reflected in adjustment of the $AdS$ radius. While the area functional is eight-dimensional, in the case of massless fundamentals, the internal space simply integrates into a prefactor holding its volume. The dynamical problem of solving for the embedding of the surfaces it equivalent to those discussed in previous subsections of $AdS_4$ and $AdS_4$-BH. The only difference comes from the overall prefactor of the area functional, which depends on the number of flavors. Explicitly, the prefactors of the surface functionals, \emph{e.g.}, in (\ref{eq:strip_EW_zs}) map to
\begin{align}
   \frac{L_y}{4 G_N^{(4)}} \rightarrow \frac{L_y}{4 G_N^{(10)}} \frac{V_6 q^2}{b^6} e^{-2\phi} = \frac{L_y}{4 G_N^{(10)}} \frac{8 \pi^3 k^2}{3(5-4b)^2(b-2)^2} \ .
\end{align}
Here $b$ is a quantity depending on the numbers of flavors, taking values from $1$ to $5/4$ as $N_f$ changes between $0$ to $\infty$: the deviation from unity is the mass anomalous dimension of the fundamentals \cite{Jokela:2013qya,Balasubramanian:2018qqx}. The constant $k$ is the Chern-Simons level and $\phi$ is a constant dilaton that we replaced for in terms of flavors.
Thus, all the previous plots apply in this case too as we omitted a constant prefactor (which here depends on the numbers of flavors).

The background with massive fundamentals, on the contrary, is very interesting since the entanglement surface explores the intrinsic space in a non-trivial manner. In this case there is an intrinsic scale in the system, there are several thermodynamic as well as information theoretic quantities that show non-monotonous behavior. For example, in \cite{Balasubramanian:2018qqx} it was demonstrated that the mutual information is non-monotonic. This is easiest to understand through the non-trivial behavior of the critical separation of the strips to undergo a phase transition between connected and disconnected phases; see figure 4 in \cite{Balasubramanian:2018qqx}. Both UV and IR in this theory possess conformal symmetry and the critical $s/l$ is therefore set by the conformal value $\varphi^{-1}=(\sqrt 5-1)/2$ (golden ratio). At intermediate energy scales the conformal symmetry is broken and the critical separation to the width ratio $s/l$ is larger in comparison to $\varphi^{-1}$. In other words, the information is more non-locally shared away from the conformal fixed points.

\begin{figure}
   \centering
   \includegraphics[width=0.6\textwidth]{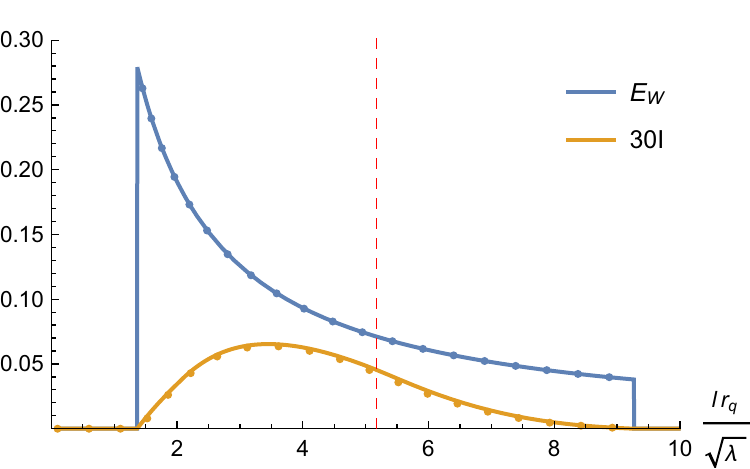}
   \caption{The information measures for the massive ABJM for $\hat\epsilon\equiv (3/4)(N_f/k)= 0.1$ \cite{Balasubramanian:2018qqx}. The dots are numerical results, while the continuous curves represent our analytical results, see the corresponding formulas in Appendix~\ref{app:abjm}. The dashed vertical line corresponds to the mass scale $r_q/(2\pi\alpha')$ of the fundamental matter, {\emph{i.e.}}, the width of a single strip just touching the bulk position $x=1$ where the sources stop contributing, see \cite{Bea:2013jxa,Balasubramanian:2018qqx}. Notice that due to chosen small $\hat\epsilon$ we needed to multiply $I$ by a large number to make it visible.}
   \label{fig:massive_EW}
\end{figure}

Furthermore, this has an interesting implication. If one chooses $s/l$ judiciously, the mutual information can be zero both in the UV and IR regions and non-vanishing only at intermediate scales.
Per inequality (\ref{eq:ineq}), we therefore expect the entanglement wedge cross section to show similar non-trivial behavior. Indeed, the $E_W$ also shares the same characteristics, see Fig.~\ref{fig:massive_EW}. This is striking behavior which is absent in theories with conformal symmetry. An entangled system of two strips at fixed $s/l$ sharing mutual information can undergo a purification just due expansion (reduction) since $E_W\to 0$ as the strip widths increase (decrease). We note that in Fig.~\ref{fig:massive_EW} we have only shown numerical results in the case when the numbers of flavors is small relative to the Chern-Simons level.  We have also included the analytic results that one can obtain in this limit (see Appendix~\ref{app:abjm}) \cite{Bea:2016ekp,Balasubramanian:2018qqx} which very nicely match the numerics. When the numbers of flavors increase, the numerical results are qualitatively the same to those in Fig.~\ref{fig:massive_EW}.

\section{Spheres} \label{sec:annulus}

In this and in the following section we will be content in discussions at zero temperature $T=0$. We therefore do not break the discussion in separate subsections as in Sec.~\ref{sec:strips}. We note that our computations can be generalized to, {\emph{e.g.}}, geometries which accommodate black holes in the interior or to confining geometries, but then one needs to resort to numerics. Here we focus on obtaining analytic results for $E_W$.

\begin{figure}
   \centering
   \includegraphics[width=0.8\textwidth]{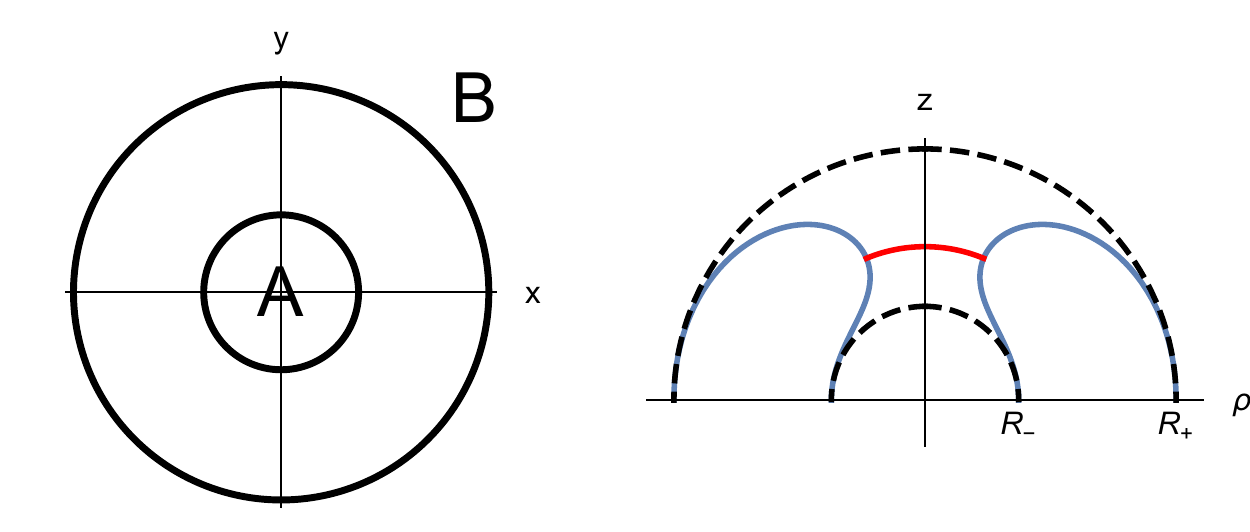}
   \caption{Left: entanglement regions in the annulus case are bounded by two concentric spheres. First region is the interior of the smaller sphere and the other region is the exterior of the larger sphere. Right: two competing minimal surfaces corresponding to the annulus. Disconnected configuration is composed of two half spheres and connected phase is a deformed half torus. The red line denotes the surface corresponding to the entanglement wedge cross section.}
   \label{fig:annulus_sketch}
\end{figure}

Let us consider a situation where the boundaries of the entanglement regions are concentric spheres with different radii, $R_\pm$ such that $R_- < R_+$ (see Fig.~\ref{fig:annulus_sketch}). In this section we mostly follow \cite{Fonda:2014cca}.
We adopt spherical coordinates on the boundary. Thus the bulk metric takes the form
\begin{align}
   g = \frac{L_{AdS}^2}{z^2} \left( -dt^2+ dz^2+ d\rho^2+ \rho^2 d\Omega_{d-2}^2 \right) \ .
\end{align}
Before writing down the area functional to be minimized, it is very useful to transform to coordinates in which the dilatation symmetry of the system is explicit. Dilatations are generated by the Killing vector $k_D=(t,z,\rho,\vec{0})/L_{AdS}$. Parameterizing the integral curves of this isometry in these coordinates are such that all $(t,z,\rho)$ are multiplied by some $e^{\lambda/L_{AdS}}$. Now we can solve for the coordinates in which this symmetry looks like a translation in only one of the new coordinates. That is, we want coordinates in which $k_D=(0,0,1,\vec{0})$. This coordinate transformation is given by
\begin{align}
   (t,z,\rho) \mapsto (\tilde t, \tilde z, u) = \left( \frac{t}{\rho}, \frac{z}{\rho}, \log\frac{\rho}{L_{AdS}} \right) \label{eq:crd_transformation} \ .
\end{align}
Now the bulk metric reads
\begin{align}
   g = \frac{L_{AdS}^2}{\tilde z^2} \left( -d\tilde t^2-2\tilde t d\tilde t du+d\tilde z^2+2\tilde z d\tilde z du+\left(-\tilde t^2+\tilde z^2+1\right)du^2+ d\Omega_{d-2}^2 \right) \ ,
\end{align}
in which form shift symmetry in $u$ is manifest. We consider a minimal surface on a time slice parameterized as $\tilde z=\tilde z(u)$. The area functional to minimize is
\begin{align}
   A = L_{AdS}^{d-1} Vol(\mathbb{S}^{d-2}) \int \frac{1}{\tilde z^{d-1}} \sqrt{1+(\tilde z+\tilde z')^2} du \ . \label{eq:annulus_area_zt}
\end{align}
The above coordinate transformation is highly useful since now the existence and form of a first integral is explicit. This conserved quantity is found through Legendre transform and reads:
\begin{align}
   -\frac{1+\tilde z(\tilde z +\tilde z')}{\tilde z^{d-1}\sqrt{1+(\tilde z + \tilde z')^2}} = \frac{1}{\sqrt K} \ ,
\end{align}
for $K>0$. The above expression can be solved for $\tilde z'(u)$,
\begin{align}
   \tilde z' = -\frac{1+\tilde z^2}{\tilde z} \left( 1\pm \frac{\tilde z^{d-2}}{\sqrt{K(1+\tilde z^2)-\tilde z^{2(d-1)}}} \right)^{-1} \ .
\end{align}
Upon integration this becomes
\begin{align}
   \int_{\log R_\pm/L_{AdS}}^{\log\rho/L_{AdS}} du = -\int_0^{\tilde z} \frac{\lambda}{1+\lambda^2} \left( 1\pm \frac{\lambda^{d-2}}{\sqrt{K(1+\lambda^2)-\lambda^{2(d-1)}}} \right) d\lambda \ ,
\end{align}
where the $R_\pm$ in the lower limit of integration expresses the boundary condition that this branch of the minimal surface should end at $\rho=R_\pm$ at the boundary. The $\tilde z$ variable takes values in range $0\le \tilde z \le \tilde z_m$, where $\tilde z_m$ is the first positive root of the square root in the expression above. In the first few lowest dimensions one can solve for $\tilde z_m$ explicitly, for example, in $d=3$ we have $\tilde z_m^2 = (K+\sqrt{K(K+4)})/2$. The above equation implies that the  branches of the solution, determined by the constant of integration $K$, can be written as
\begin{align}
   \begin{cases}
      \rho = R_-\exp(-f_{-,K}(\tilde z)) \\
      \rho = R_+\exp(-f_{+,K}(\tilde z))
   \end{cases} \label{eq:annulus_rho_sol} \ ,
\end{align}
where
\begin{align}
   f_{\pm,K}(\tilde z) = \int_0^{\tilde z} \frac{\lambda}{1+\lambda^2} \left( 1\pm \frac{\lambda^{d-2}}{\sqrt{K(1+\lambda^2)-\lambda^{2(d-1)}}} \right) d\lambda \ .
\end{align}

Note that the radii $R_\pm$ in (\ref{eq:annulus_rho_sol}) are not independent. We have to glue both branches together at $\tilde z = \tilde z_m$ which gives us the following relation between $R_-$ and $R_+$:
\be\label{eq:Rmpvszm}
   -\log\frac{R_-}{R_+} = \int_0^{\tilde z_m} \frac{2\lambda^{d-1}}{(1+\lambda^2)\sqrt{K(1+\lambda^2)-\lambda^{2(d-1)}}} d\lambda \ .
\ee
For each possible $R_-/R_+$, there are two different values of $K$ corresponding to two local minima of the bulk surface. It turns out that out of these two surfaces, the one with a lower value of $K$ has always the smaller area \cite{Fonda:2014cca}. One should also note that there is a positive minimum for $R_-/R_+$ as a function of $K$. It means that connected surfaces only exist if the inner radius is not too small compared to the outer radius. If a connected surface does not exist, then the global minimum is easily found to be a union of disjoint coordinate half-spheres in the bulk with radii $R_-$ and $R_+$.

As our primary interest is $d=3$, we will quote the explicit form of the above expression here\footnote{Notice that the expression (\ref{eq:annulus_profile}) differs from the one in \cite{Fonda:2014cca}, but the difference does not alter their later analysis, since (\ref{eq:annulus_area_fp}) only depends on the derivative of $f_{\pm,K}(\tilde z)$.}
\begin{gather} \label{eq:Rmpvszm_d3}
  f_{\pm,K}(\tilde z) = \frac{1}{2}\log\left(1+\tilde z^2\right) \pm \frac{\kappa}{\sqrt{2\kappa^2-1}} \Bigg( \mathbb{F}\left(\arcsin\left(\frac{\tilde z}{\tilde z_m}\right),\frac{\kappa^2}{\kappa^2-1}\right) \nonumber \\
  + \Pi\left( \frac{2\kappa^2-1}{\kappa^2-1}, \arcsin\left( \frac{\tilde z}{\tilde z_m}, \frac{\kappa^2}{\kappa^2-1} \right), \frac{\kappa^2}{\kappa^2-1} \right) \Bigg) \label{eq:annulus_profile} \\
   -\log\frac{R_-}{R_+}\Bigg|_{d=3} = \frac{\sqrt{\tilde z_m^2+1}}{\tilde z_m} \left( \Pi(-\tilde z_m^2,-\tilde z_m^2-1)-\mathbb{K}(-\tilde z_m^2-1) \right) \ ,
\end{gather}
where $\mathbb{K}(a)$ and $\Pi(a,b)$ are the complete elliptic integrals of first and third kinds, respectively. $\mathbb{F}(a,b)$ and $\Pi(a,b,c)$ are the incomplete versions of the same elliptic integrals.

Before finding the actual entanglement wedge cross section we must compute the entanglement entropy in order to determine the correct phase for the system. To find the minimal area, we substitute the solution (\ref{eq:annulus_rho_sol}) to the area functional. This results in
\begin{align}
   \frac{A_\pm}{L_{AdS}^{d-1} Vol(\mathbb{S}^{d-2})} &= \int \frac{1}{\tilde z^{d-1}} \sqrt{1+ \left( \tilde z+\frac{\rho_\pm}{\rho_\pm'} \right)^2 } \frac{\rho_\pm'}{\rho_\pm} d\tilde z \\
                                                     &= \int \frac{1}{\tilde z^{d-1}} \sqrt{1+ \left( \tilde z - \frac{1}{f_{\pm,K}'(\tilde z)} \right) } f'_{\pm,K}(\tilde z) d\tilde z \label{eq:annulus_area_fp} \\
   &= \int_0^{\tilde z_m} \frac{1}{\tilde z^{d-1}\sqrt{1+\tilde z^2 - \tilde z^{2(d-1)}/K}} d\tilde z \ .
\end{align}
The integrals are the same for both branches. Note though, that as usual, there is an UV-divergence at $\tilde z=0$ we must subtract. The UV-cutoff introduces dependence on the boundary radius, making the integrals of different branches unequal. The divergent contribution we subtract depends on $d$:
\begin{align}
   \frac{A_\pm^\text{div}}{L_{AdS}^{d-1} Vol(\mathbb{S}^{d-2})} = \int_{\epsilon/R_\pm}^{\tilde z_m} d\tilde z \left( \frac{1}{\tilde z^{d-1}} + \frac{C_{d-3}}{\tilde z^{d-3}} + \frac{C_{d-5}}{\tilde z^{d-5}} + \ldots + \begin{cases} \frac{C_{-1}}{\tilde z} + \mathcal{O}(\tilde z)& \text{odd } d \\ C_0 +\mathcal{O}(\tilde z^2) & \text{even } d \end{cases} \right) \ ,
\end{align}
where $\epsilon$ is the UV-cutoff and $R_\pm$ is the boundary value of $\rho$ in the corresponding branch. We need to subtract two of these divergences since there are two branches in the solution. For completeness, we will again state explicitly the case $d=3$. In this dimension, the connected phase of the annulus has the following area
\begin{align}
   \frac{A_\text{connected}}{2\pi L_{AdS}^2}\Bigg|_{d=3} &= \frac{R_-+R_+}{\epsilon} - \frac{2}{\tilde z_m} + 2\int_0^{\tilde z_m} \frac{1}{\tilde z^2} \left( \frac{1}{\sqrt{1+\tilde z^2-\tilde z^4/K}} - 1 \right) d\tilde z \\
   &= \frac{R_-+R_+}{\epsilon} + \frac{2}{\tilde z_m} \left( \mathbb{K}(-\tilde z_m^2-1) - \mathbb{E}(-\tilde z_m^2-1) \right) \ ,
\end{align}
where $\mathbb{E}$ is the complete elliptic integral of the second kind.

The disconnected phase is simply a pair of half spheres, $R_\pm^2=\rho^2+z^2$. Their area is given by
\begin{align}
   \frac{A_\text{disconnected}}{Vol(\mathbb{S}^{d-2}) L_{AdS}^{d-1}} = \int_{\epsilon/R_-}^1 \frac{(1-y^2)^{(d-3)/2}}{y^{d-1}} + \int_{\epsilon/R_+}^1 \frac{(1-y^2)^{(d-3)/2}}{y^{d-1}} \ .
\end{align}
This integral has the same UV-divergences as the connected configuration discussed previously. In $d=3$, explicitly,
\begin{align}
   \frac{A_\text{disconnected}}{2\pi L_{AdS}^2}\Bigg|_{d=3} = \frac{R_- + R_+}{\epsilon} - 2 \ .
\end{align}

The phase of the system can be determined by the sign of the following finite quantity
\begin{align}
   \Delta A &= A_\text{disconnected} - A_\text{connected} \\
   &= Vol(\mathbb{S}^{d-2}) L_{AdS}^{d-1} \left( \int_0^\infty \frac{1}{\tilde z^{d-1}\sqrt{1+\tilde z^2}} d\tilde z - \int_0^{\tilde z_m} \frac{1}{\tilde z^{d-1}\sqrt{1+\tilde z^2 - \tilde z^{2(d-1)}/K}} d\tilde z \right) \ . \label{eq:annulus_delta_A}
\end{align}
Both integrals have UV-divergences but in the subtraction, they cancel each other out. Again, we write the $d=3$ case explicitly
\begin{align}
   \Delta A|_{d=3} = 4\pi L_{AdS}^2 \left( \frac{\mathbb{E}(-\tilde z_m^2-1)-\mathbb{K}(-\tilde z_m^2-1)}{\tilde z_m} - 1 \right) \ .
\end{align}

Note that $\Delta A$ is closely related to the mutual information $I(A,B)$. When $\Delta A$ is positive, that is, $S(AB)$ is in its connected phase, we have
\begin{align}
  I(A,B) = \frac{1}{4 G_N^{(d+1)}} \Delta A \ .
\end{align}

\begin{figure}
   \centering
   \includegraphics[width=0.6\textwidth]{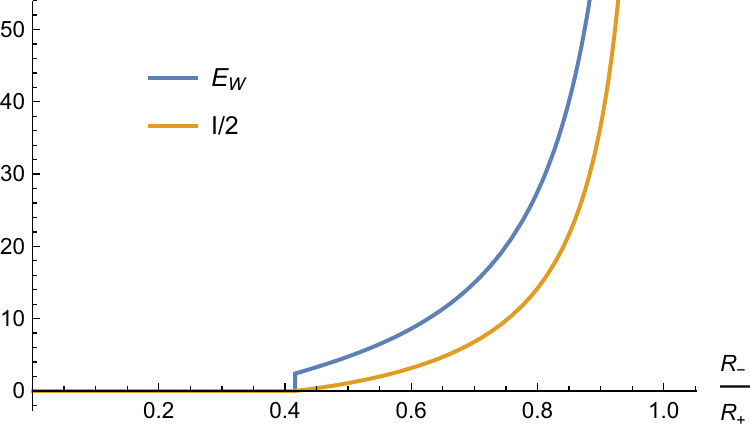}
   \caption{Entanglement wedge cross section and mutual information for an annulus in $AdS_4$. We have omitted the common prefactor $L_{AdS}^{2}/4G_N^{(4)}$ from both quantities.}
   \label{fig:annulus_EW}
\end{figure}

In order to find the entanglement wedge cross section, we must find the minimal surface which splits the wedge in regions associated with $A$ and $B$. Since the area of this surface must be minimized, the surface must be a section of a coordinate half-sphere with a boundary radius $R$ such that $R_-<R<R_+$. The area of this surface can be written as
\begin{align}
   Vol(\mathbb{S}^{d-2}) L_{AdS}^{d-1} \int_{\tilde z_*}^\infty \frac{d\tilde z}{\tilde z^{d-1} \sqrt{1+\tilde z^2}} \ ,
\end{align}
where the upper limit corresponds to $\rho=0$ and the lower limit is determined by the point where this surface meets the minimal surface of the connected configuration of $S(AB)$. This area is a monotonically decreasing function of $\tilde z_*$, meaning that the area is minimized when $\tilde z_*=\tilde z_m$. Thus, assuming $R_\pm$ are such that the connected configuration exists and (\ref{eq:annulus_delta_A}) is positive, the entanglement wedge cross section is
\begin{align}
   E_W = \frac{Vol(\mathbb{S}^{d-2}) L_{AdS}^{d-1}}{4 G_N^{(d+1)}} \int_{\tilde z_m}^\infty \frac{d\tilde z}{\tilde z^{d-1} \sqrt{1+\tilde z^2}} \ .
\end{align}
These integrals are easy to calculate and as previously, we explicitly quote the result for $d=3$:
\begin{align}
   E_W = \frac{2\pi L_{AdS}^{d-1}}{4 G_N^{(d+1)}} \left( \sqrt{1+\frac{1}{\tilde z_m^2}} -1 \right)  \ .
\end{align}
The final result for $E_W=E_W(R_-/R_+)$ is obtained via substituting $\tilde z_m=\tilde z_m(R_-/R_+)$ from (\ref{eq:Rmpvszm}) in general or from (\ref{eq:Rmpvszm_d3}) in $d=3$. In Fig.~\ref{fig:annulus_EW} we plot $E_W$ and $I/2$ as functions of $R_-/R_+$ in $d=3$.

\section{Creases} \label{sec:slice}

In this section we will consider surfaces that are not smooth. The simplest configurations to consider are entangling regions that have a string-like singularity, those that are denoted by  $k\times \mathbb{R}^{d-3}$ in \cite{Myers:2012vs}; see figure 1c in that paper for visualization. These configurations are interesting to study as the corner contribution is universal and  carries a physical interpretation: in \cite{Myers:2012vs,Bueno:2015rda,Bueno:2015xda,Faulkner:2015csl} it was shown that they can be related to the central charge of the UV conformal field theory.

Let us thus follow \cite{Myers:2012vs} and prepare the calculation of the $E_W$ for two entangling creases in generic dimension. We start by writing the bulk metric as follows
\begin{align}
  g = \frac{L_{AdS}^2}{z^2} \left( -dt^2+ dz^2+ d\rho^2+ \rho^2 d\phi^2+ d\vec{x}_{d-3}^2 \right) \ ,
\end{align}
where $\rho\geq 0$, $\phi\in(0,2\pi)$ and $\vec{x}_{d-3}$ is the space contained in the singular locus of the crease. With the same coordinate transformation we used for the spheres, eq. (\ref{eq:crd_transformation}), the metric takes the following more useful form

\begin{align}
  g = \frac{L_{AdS}^2}{z^2} \left( -d\tilde t^2- 2\tilde t d\tilde t du+ d\tilde z^2+ 2\tilde z d\tilde z du + \left( -\tilde t^2 + \tilde z^2 + 1 \right) du^2 + d\phi^2 + \frac{d\vec{x}_{d-3}^2}{L_{AdS}^2 e^{2u}}\right) \ .
\end{align}
The minimal surface we are looking for lies at the time slice $\tilde t=0$ and is parameterized by $\phi=\phi(\tilde z)$. The functional giving the area of this surface is
\begin{align}
   A=V L_{AdS}^{d-1} \int \frac{d\rho}{\rho^{d-2}} \int \frac{1}{\tilde z^{d-1}} \sqrt{1+(1+\tilde z^2)\phi'(\tilde z)^2} d\tilde z \label{eq:crease_functional} \ ,
\end{align}
where $V=\int d\vec{x}_{d-3}$. The equation of motion governing $\phi(\tilde z)$ has a conserved quantity which can be written as \cite{Myers:2012vs}
\begin{gather}
   \frac{\left(1+\tilde z^2\right)^{(d-1)/2}}{\tilde z^{d-1} \sqrt{1+\left(1+\tilde z^2\right) \phi'(\tilde z)^2}} = \frac{1}{\sqrt K} \\
   \rightarrow \phi'(\tilde z) = \pm\frac{\tilde z^{d-1}}{\sqrt{\left(1+\tilde z^2\right) \left(K \left(1+\tilde z^2\right)^{d-2}-\tilde z^{2 (d-1)}\right)}} \label{eq:phi_derivative} \ .
\end{gather}
The minimal embedding is given as the integral of the above expression
\begin{align}
   \phi(\tilde z) = \int_0^{\tilde z} \frac{\lambda^{d-1}}{\sqrt{\left(1+\lambda^2\right) \left(K \left(1+\lambda^2\right)^{d-2}-\lambda^{2 (d-1)}\right)}} d\lambda \ .
\end{align}
A qualitative difference with the sphere section is that this time, we have only one extremal surface for each crease. Again, $\tilde z$ takes values between zero (corresponds to the boundary) and some $\tilde z_m$ which is given by the first positive root of the expression in the square root in (\ref{eq:phi_derivative}).

The opening angle of the crease is related to $\tilde z_m$ by
\begin{align}
  \Omega = 2\phi(\tilde z_m) \label{eq:crease_omega} \ .
\end{align}
This function $\Omega=\Omega(\tilde z_m)$ is such that it maps positive numbers monotonically to the range $(0,\pi)$.

The area of a crease is found by plugging (\ref{eq:phi_derivative}) into (\ref{eq:crease_functional})
\begin{align}
   \frac{A}{2 V L_{AdS}^{d-1}} &=\int_\delta^R \frac{d\rho}{\rho^{d-2}} \int_{\epsilon/\rho}^{\tilde z_m} \frac{1}{\tilde z^{d-1}}\sqrt{1-\frac{1}{1-K \left(1+\tilde z^2\right)^{d-2} \tilde z^{2 (1-d)}}} d\tilde z \\
   &=\int_\delta^R \frac{d\rho}{\rho^{d-2}} \Bigg[ \frac{(\epsilon/\rho)^{2-d}}{d-2}-\frac{\tilde z_m^{2-d}}{d-2} \nonumber \\
   &\quad+\int_0^{\tilde z_m} \frac{1}{\tilde z^{d-1}} \left(\sqrt{1-\frac{1}{1-K \left(1+\tilde z^2\right)^{d-2} \tilde z^{2(1-d)}}} -1\right) d\tilde z \Bigg] \\
   &= \frac{R-\delta}{(d-2)\epsilon^{d-2}} + \frac{1}{(d-3)\delta^{d-3}} \mathcal{F}_d(\tilde z_m) + \mathcal{O}\left( \frac{1}{R^{d-3}} \right) \ ,
\end{align}
where
\begin{align}
   \mathcal{F}(\tilde z_m) &= \int_0^{\tilde z_m} \frac{1}{\tilde z^{d-1}} \left(\sqrt{1-\frac{1}{1-K \left(1+\tilde z^2\right)^{d-2} \tilde z^{2(1-d)}}}-1\right) d\tilde z-\frac{\tilde z_m^{2-d}}{d-2} \ .
\end{align}
Although now written as a function of $\tilde z_m$, we consider $A$ and $\mathcal{F}$ functions of $\Omega$ through inverting (\ref{eq:crease_omega}), $\tilde z_m = \tilde z_m(\Omega)$. These functions are only defined on $\Omega\in(0,\pi)$. Since we are working on a pure state, observables will have the symmetry $\Omega\to 2\pi-\Omega$, for example, we define $A(\pi<\Omega<2\pi)=A(2\pi-\Omega)$. Entanglement entropy is then simply $S=A/4G_N^{(d+1)}$,
\begin{align}
   S = \frac{V L_{AdS}^{d-1}}{2 G_N^{(d+1)}} \left( \frac{R-\delta}{(d-2)\epsilon^{d-2}} + \frac{\mathcal{F}_d(\tilde z_m)}{(d-3)\delta^{d-3}} \right) \ .
\end{align}
The divergence structure is modified in $d=3$ to
\begin{align}
  S^{(d=3)} = \frac{L_{AdS}^2}{2 G_N^{(4)}} \left( \frac{R}{\epsilon} - \frac{\delta}{\epsilon} + \log\left(\frac{R}{\delta}\right) \mathcal{F}_3(\tilde z_m) \right) .
\end{align}
Notice that the cutoff $\epsilon$ is the usual UV-divergence originating from the conformal boundary $z=\epsilon$ and the other divergence, $\delta$, cutting off the singular locus is only present due to the singularity of the surface and is associated with the IR, as can be seen through mapping the singularity to the IR singularity of an infinite slab \cite{Bueno:2015xda}.\footnote{Notice, that a crease on $\mathbb{R}^d$ maps to a slab on flat $\mathbb{R}^d$ only when the opening angle is small. However, without any restrictions on the opening angle, the crease maps to a slab on $\mathbb{R}^{d-2}\times S^2$ in such a way that the cutoffs $\delta$ and $R$ are equivalent to cutting off the infinite length of the slabs.}

\begin{figure}
   \centering
   \includegraphics[width=0.8\textwidth]{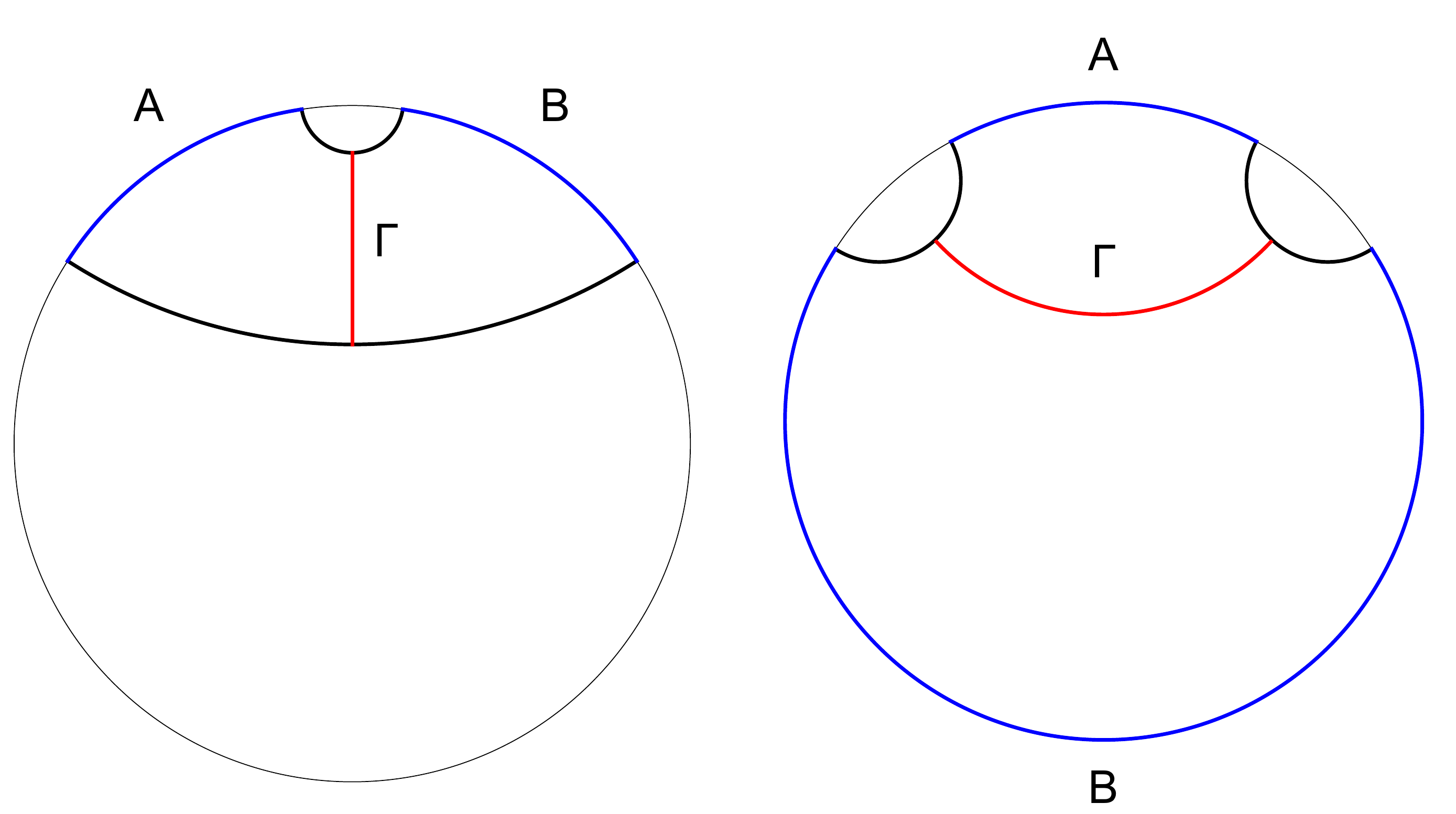}
   \caption{Connected crease configurations for $S(AB)$. Associated wedge cross sections are marked in red. The left configuration corresponds to the case of symmetric creases, $\Omega_1=\Omega_2=\Omega$, which is the one where we present our explicit calculations. The right configuration represents the more general situation where the creases need not have the same size.}
   \label{fig:crease_configs}
\end{figure}
Consider two creases with no non-zero overlap and opening angles $\Omega_1$ and $\Omega_2$, separated by $\varphi$. The corresponding mutual information is (see also \cite{Mozaffar:2015xue})
\begin{align}
   I &= S(\Omega_1) + S(\Omega_2) - S(\Omega_1+\Omega_2+\varphi) - S(\varphi) \\
   &= \frac{V L_{AdS}^{d-1}}{2 G_N^{(d+1)}} \frac{1}{(d-3)\delta^{d-3}} \left( \mathcal{F}_d(\Omega_1) + \mathcal{F}_d(\Omega_2) - \mathcal{F}_d(\Omega_1+\Omega_2+\varphi) - \mathcal{F}(\varphi) \right) \ .
\end{align}
We will now consider symmetric crease configurations ($\Omega_1=\Omega_2=\Omega$) because they have a particularly simple entanglement wedge cross section (left configuration in Fig. \ref{fig:crease_configs}). The minimal surface extends between $\tilde z=\tilde z_m(\varphi)$ and $\tilde z=\tilde z_m(2\Omega+\varphi)$ and lie along a constant $\phi$. The exact form of $E_W$ depends on the values of $\varphi$ and $2\Omega+\varphi$. First assuming $0<\varphi<2\Omega+\varphi<\pi$ we have
\begin{align}
  E_W &= \frac{V L_{AdS}^{d-1}}{4 G_N^{(d+1)}} \frac{1}{(d-3)\delta^{d-3}} \int_{\tilde z_m(\varphi)}^{\tilde z_m(2\Omega+\varphi)} \frac{d\tilde z}{\tilde z^{d-1}} \\
  &= \frac{V L_{AdS}^{d-1}}{4 G_N^{(d+1)}} \frac{1}{(d-2)(d-3)\delta^{d-3}} \left( \frac{1}{\tilde z_m(\varphi)^{d-2}} - \frac{1}{\tilde z_m(2\Omega+\varphi)^{d-2}} \right) \ .
\end{align}
On the other hand, if $0<\varphi<\pi<2\Omega+\varphi<2\pi$,
\begin{align}
  E_W &= \frac{V L_{AdS}^{d-1}}{4 G_N^{(d+1)}} \frac{1}{(d-3)\delta^{d-3}} \left( \int_{\tilde z_m(\varphi)}^\infty \frac{d\tilde z}{\tilde z^{d-1}} + \int_{\tilde z_m(2\pi-2\Omega-\varphi)}^\infty \frac{d\tilde z}{\tilde z^{d-1}} \right) \\
  &= \frac{V L_{AdS}^{d-1}}{4 G_N^{(d+1)}} \frac{1}{(d-2)(d-3)\delta^{d-3}} \left( \frac{1}{\tilde z_m(\varphi)^{d-2}} + \frac{1}{\tilde z_m(2\pi-2\Omega-\varphi)^{d-2}} \right) \ .
\end{align}
In the last case, $\pi<\varphi<2\Omega+\varphi$, we have
\begin{align}
  E_W &= \frac{V L_{AdS}^{d-1}}{4 G_N^{(d+1)}} \frac{1}{(d-3)\delta^{d-3}} \int_{\tilde z_m(2\pi-2\Omega-\varphi)}^{\tilde z_m(2\pi-\varphi)} \frac{d\tilde z}{\tilde z^{d-1}} \\
  &= \frac{V L_{AdS}^{d-1}}{4 G_N^{(d+1)}} \frac{1}{(d-2)(d-3)\delta^{d-3}} \left( \frac{1}{\tilde z_m(2\pi-2\Omega-\varphi)^{d-2}} - \frac{1}{\tilde z_m(2\pi-\varphi)^{d-2}} \right) \ .
\end{align}
Again, in $d=3$ the divergence becomes logarithmic. 

\begin{figure}
  \centering
  \includegraphics[width=\textwidth]{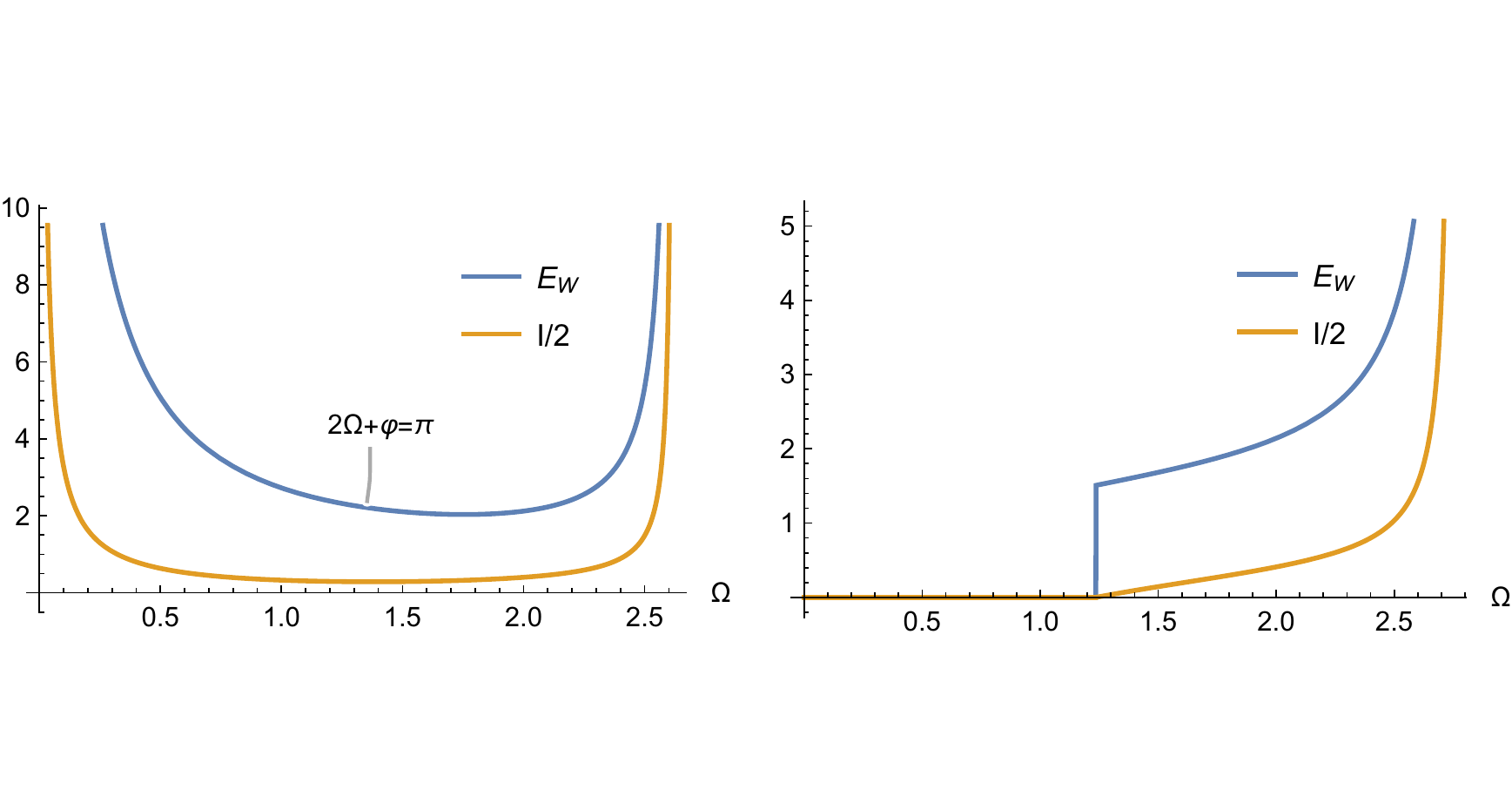}
  \caption{Left: $E_W$ and $I/2$ for a symmetric configuration $\Omega_1=\Omega_2=\Omega$ and the separation fixed to $\varphi=0.4\Omega$. For small $\Omega$, both quantities decrease as the corresponding surfaces are lowered deeper into the bulk. At some point when $2\Omega+\varphi$ starts to approach $2\pi$, the bulk surfaces are pulled back towards the conformal boundary, causing a divergence in both $E_W$ and $I/2$. Right: Crease configuration is still symmetric $\Omega_1=\Omega_2=\Omega$ but now the separation is fixed to $\varphi=\pi/4$. When $\Omega$ is too small, the entanglement wedge is disconnected and $E_W=I=0$. When increasing $\Omega$ over some point, the entanglement wedge becomes connected and both quantities become positive and increase monotonically towards a divergence which is caused by $2\Omega+\varphi$ approaching $2\pi$. In both panels we have set $d=3$.}
  \label{fig:crease_EW}
\end{figure}

In Fig.~\ref{fig:crease_EW} we have depicted the case $d=3$. Again, we find that $E_W$ is always greater than half the mutual information. In the left panel of Fig.~\ref{fig:crease_EW} we find that $E_W$ is not monotonically increasing as a function of $\Omega$, however, as we are increasing the opening angle $\Omega$, we are simultaneously scaling $\varphi$. The qualitative features of $E_W$ can be understood directly from the geometry: $E_W$ reaches its global minimum for some $2\Omega+\varphi>\pi$ (depending on $\varphi$) beyond which $E_W$ begins to grow again and eventually diverging when $2\Omega+\pi\to 2\pi$ as bulk surfaces approach the conformal boundary. In the right panel of Fig.~\ref{fig:crease_EW} we are instead keeping $\varphi$ fixed and dialing the opening angles $\Omega$. The $E_W$ has the expected behavior: for small $\Omega$ the configuration is in the disconnected phase, $E_W=0$, and at some point $E_W$ jumps to a positive value and keeps monotonically increasing due to transitioning to the ever-expanding connected phase.

We now briefly consider more general configurations (right panel in Fig.~\ref{fig:crease_configs}) which correspond to connected entanglement wedges and thus positive $E_W$. Since the opening angles of the two creases are no longer required to be equal, the cross section surface need not be flat. Still, since the cross section should have minimal area, this surface must be a subset of the same minimal bulk surface that corresponds to a single boundary crease. The cross section area is
\begin{align}
  E_W &= \frac{V L_{AdS}^{d-1}}{4 G_N^{(d+1)}} \frac{1}{(d-3)\delta^{d-3}} \Bigg( \int_{\tilde z_1}^{\tilde z_*} \frac{1}{\tilde z^{d-1}} \sqrt{1-\frac{1}{1-K \left(1+\tilde z^2\right)^{d-2} \tilde z^{2 (1-d)}}} d\tilde z \nonumber \\
      &+ \int_{\tilde z_2}^{\tilde z_*} \frac{1}{\tilde z^{d-1}} \sqrt{1-\frac{1}{1-K \left(1+\tilde z^2\right)^{d-2} \tilde z^{2 (1-d)}}} d\tilde z \Bigg) \ ,
\end{align}
where $\tilde z_*$ is the turning point of the surface in $\tilde z$ and $\tilde z_1$ ($\tilde z_2$) is the point where the left (right) branch of the minimal surface ends on the edge of the entanglement wedge. There is one constraint equation on $(\tilde z_*, \tilde z_1, \tilde z_2)$, so the parameter space is two dimensional. The problem of finding the minimum in this parameter space cannot be solved analytically, but one can efficiently find numerical solutions using gradient descent based methods.

\section{Conclusions} \label{sec:discussion}
The entanglement of purification has been the topic of many recent works. The holographic candidate for this quantity is the entanglement wedge cross section. We feel that there is still lots to establish until this relationship has been satisfactorily shown. In this paper we made important progress towards this goal.

The entanglement wedge cross section has been computed primarily for global asymptotically $AdS_3$ spacetimes, where the entangling regions sharing mutual information are two line segments. We computed $E_W$ in various other higher dimensional geometries, in Poincar\'e coordinates, where the entangling regions were slabs, concentric spheres, and creases. We also point out in Appendix~\ref{app:mapping} that the latter two can be mapped to global coordinates, so our results are then directly applicable to systems, where the regions of interest are caps and orange slices on hyperspheres, respectively.

We demonstrated that both the mutual information and the entangling wedge cross section are generically not monotonous functions of the scales in the problem if the conformal invariance is broken in the background. In particular, in the large-$N$ limit, the $E_W$ can feature discontinuous jumps upwards when the system size is taken larger.

\vspace{1cm}
\paragraph{Acknowledgments}

We thank Esko Keski-Vakkuri for discussions. A.~P. acknowledges support from the Vilho, Yrj\"o and Kalle V\"ais\"al\"a Foundation of the Finnish Academy of Science and Letters.

\appendix

\section{Mapping results to global coordinates} \label{app:mapping}

Let us now write down a coordinate transformation between the Poincar\'e and global coordinate charts of $AdS_{d+1}$. This will enable us to transfer some of our Poincar\'e patch results derived in previous sections to associated problems in global $AdS_{d+1}$.

We denote the Poincar\'e coordinates by ($t,z,x_i,\dots,x_{d-1}$) and global coordinates by ($\tau,r,\theta,\phi_i,\dots,\phi_{d-2}$). The mapping between the charts is
\begin{align}
   t &= \frac{L_{AdS} \sqrt{(r/L_{AdS})^2+1}\sin(\tau/L_{AdS})}{\sqrt{(r/L_{AdS})^2+1}\cos(\tau/L_{AdS})+(r/L_{AdS})\cos\theta} \\
   z &= \frac{L_{AdS}}{\sqrt{(r/L_{AdS})^2+1}\cos(\tau/L_{AdS})+(r/L_{AdS})\cos\theta} \\
   x_i &= \frac{r \sin\theta \cos\phi_i \prod_{j=1}^{i-1}\sin\phi_j}{\sqrt{(r/L_{AdS})^2+1}\cos(\tau/L_{AdS})+(r/L_{AdS})\cos\theta}\qquad i=1,\dots,d-2 \\
   x_{d-1} &= \frac{r \sin\theta \prod_{j=1}^{d-2}\sin\phi_j}{\sqrt{(r/L_{AdS})^2+1}\cos(\tau/L_{AdS})+(r/L_{AdS})\cos\theta} \ .
\end{align}
The above mapping brings the Poincar\'e metric to
\begin{align}
  g = -\left( 1+\frac{r^2}{L_{AdS}^2} \right) dt^2 + \frac{dr^2}{1+\frac{r^2}{L_{AdS}^2}} + r^2 d\Omega_{d-1}^2 \ .
\end{align}
In order to map results to global $AdS$, it is important to understand how the boundaries map to each other since this tells us which entanglement regions correspond to each other on different sides of the mapping.

We first consider spheres of Sec.~\ref{sec:annulus}. The entanglement regions are defined by $\sqrt{\sum_i^{d-1} x_i^2}=R$. Translating this constraint to global coordinates gives
\begin{align}
  \sum_i^{d-1} x_i^2 = L_{AdS} \tan\frac{\theta}{2} \ ,
\end{align}
implying that a sphere in Poincar\'e coordinates corresponds to an entanglement region defined by $\theta=\theta_0$ for some $\theta_0\in(0,\pi)$. We will call these regions polar caps. All calculations of Sec.~\ref{sec:annulus} can be readily applied to find the entanglement entropy of a polar cap, $S(\theta_0)$, or the mutual information $I(\theta_A,\theta_B)$ and entanglement wedge cross section $E_W(\theta_A,\theta_B)$ of two concentric polar cap regions simply by substituting $R=L_{AdS} \tan(\theta_0/2)$ into appropriate formulas. The relation between $R$ and $\theta_0$ implies that disks of radii $R<L_{AdS}$ are mapped to the northern hemisphere $\theta<\pi/2$ and radii $R>L_{AdS}$ are mapped to the southern hemisphere of the compact target space. Bulk surfaces can be converted by using
\begin{align}
  \tilde z = L_{AdS}\frac{\csc\theta}{r} \ .
\end{align}

Similar substitutions can be done for the crease of Sec.~\ref{sec:slice} by noting that if one takes the polar coordinates to correspond to $x_{d-2}$ and $x_{d-1}$, then we have the boundary coordinate relation $\phi=\phi_{d-2}$. The Poincar\'e patch entanglement region defined by $-\Omega/2\leq \phi\leq\Omega/2$ transforms to $-\Omega/2\leq \phi_{d-2}\leq\Omega/2$ in the global patch.

\section{Analytic formulas: massive ABJM}\label{app:abjm}
In this appendix we simply list the analytic formulas needed for generating the Fig.~\ref{fig:massive_EW}. We refer the reader to Section 3.1 of \cite{Balasubramanian:2018qqx} for detailed derivations of the entanglement entropy $S(x_*)$ as a function of strip width $l(x_*)$, where $x_*$ is the tip position of the hanging strip in radial coordinate $x$. The reversion between $x_*$ and $l$ is involved, which leads to quite convoluted formulas. The results for the entanglement wedge cross section and the mutual information, however, simply follows from the construction of $S(l)$ with meticulously paying attention on reverting $x_* \leftrightarrow l$.

The relevant formulas are as follows:
\begin{align}
   E_W &= \frac{L_y V_6}{4 G_N^{(10)}} \int_{x_*^{(1)}}^{x_*^{(2)}} \sqrt{G(x) H(x)} dx \label{eq:ABJMEW}\\
   S(x_*) &= \frac{L_y V_6}{2 G_N^{(10)}} \int_{x_*}^\infty \left( \frac{\sqrt{G(x)}H(x)}{\sqrt{H(x)-H(x_*)}} - \sqrt{G_\infty H_\infty}x^{\frac{1}{b}-1}\right) dx \nonumber \\
   &+ \frac{L_y V_6}{2 G_N^{(10)}} b \sqrt{G_\infty H_\infty} \left( x_{max}^\frac{1}{b} - x_*^\frac{1}{b} \right) \\
   l(x_*) &= 2 \int_{x_*}^\infty \frac{G(x)}{\sqrt{\frac{H(x)}{H(x_*)}}-1} dx \ ,
\end{align}
where
\begin{align}
   G(x) &= \frac{L_{ABM}^4}{r_q^2}\frac{1}{x^4} \left( 1+\hat\epsilon\frac{25-28x^2-189x^4+140x^4\log x}{280x^4} + \mathcal{O}(\hat\epsilon^2) \right)  \\
   H(x) &= \frac{k^4r_q^4L_{ABJM}^4}{16} x^4 \left( 1-\hat\epsilon\frac{5+84x^2-329x^4+140x^4\log x}{140x^4} + \mathcal{O}(\hat\epsilon^2)\right)
\end{align}
plugged back in to (\ref{eq:ABJMEW}) casts $E_W$ in the following explicit form
\begin{align}
   \frac{E_W}{k^2 L_{ABJM}^4 r_q} &= \frac{1}{4} \left(x_*^{(2)}-x_*^{(1)}\right) + \frac{\hat\epsilon}{2240}
   \begin{cases}
      A(x_*^{(1)},x_*^{(2)}) & ,\ 1\leq x_*^{(1)} < x_*^{(2)} \\
      B(x_*^{(1)},x_*^{(2)}) & ,\ 0<x_*^{(1)}<1\leq x_*^{(2)} \\
      C(x_*^{(1)},x_*^{(2)}) & ,\ 0<x_*^{(1)}<x_*^{(2)}<1 \ ,
   \end{cases}
\end{align}
where
\bea
 A & = &\frac{5}{\left(x_*^{(1)}\right)^3}-609 x_*^{(1)}-\frac{196}{x_*^{(1)}}+140 x_*^{(1)} \log x_*^{(1)}-\frac{5}{\left(x_*^{(2)}\right)^3}+609 x_*^{(2)}+\frac{196}{x_*^{(2)}}-140 x_*^{(2)} \log x_*^{(2)} \nonumber\\
 B & = & -16 \left[\left(x_*^{(1)}\right)^4+14 x_*^{(1)}+35\right]-\frac{5}{\left(x_*^{(2)}\right)^3}+609 x_*^{(2)}+\frac{196}{x_*^{(2)}}-140 x_*^{(2)} \log x_*^{(2)}  \nonumber\\
 C & = & 16\left(\left(x_*^{(2)}\right)^4-\left(x_*^{(1)}\right)^4+14\left[x_*^{(2)}-x_*^{(2)}\right]\right) \ .
\eea
For the sake of compactness, in the above formulas we have defined $x_*^{(1)}=x_*(s)$ and $x_*^{(2)}=x_*(2l+s)$, where $l$ is the width of the slabs and $s$ is the distance separating them.

\end{document}